# Natural Language Processing (NLP) tools for Pharmaceutical Manufacturing Information Extraction from Patents

Diego Alvarado-Maldonado*, Blair Johnston and Cameron J. Brown

Abundant and diverse data on medicines manufacturing and other lifecycle components has been made easily accessible in the last decades. However, a significant proportion of this information is characterised by not being tabulated and usable for machine learning purposes. Thus, natural language processing (NLP) tools have been used to build databases in domains such as biomedical and chemical to address this limitation. This has allowed the development of artificial intelligence (AI) applications, which have improved drug discovery and treatments. In the pharmaceutical manufacturing context, some initiatives and datasets for primary processing can be found, but the manufacturing of drug products is an area which is still lacking, to the best of our knowledge. This works aims to explore and adapt NLP tools used in other domains to extract information on both primary and secondary manufacturing, employing patents as the main source of data. Thus, two independent, but complementary, models were developed comprising 1) a method to select fragments of text that contain manufacturing data, and 2) a named entity recognition system that enables extracting information on operations, materials, and conditions of a process. For the first model, the identification of relevant sections was achieved using an unsupervised approach combining Latent Dirichlet Allocation and k-Means clustering. The performance of this model measured as a Cohen's kappa between model output and manual revision was higher than 90%. NER model consisted of a deep neural network, and an f1-score micro average of 84.2% was obtained which is comparable to other works. Some considerations for these tools to be used in data extraction are discussed throughout this document.

## 1. Introduction

Over time, information on the design and development of medicines have undergone an intensive process of democratization. During this process, a vast amount of data has been made accessible by the private sector, regulatory agencies and academic institutions. As a result, a massive amount of information is available to be consulted in digital sources[1]. The available information covers all pharmaceutical product lifecycle, ranging from design and development of new chemical entities up to pharmacovigilance and safety aspects[1–3]. Consequently, pharmaceutical information can be said to be varied and abundant. However, another important feature is that much of this information is unstructured[1,3]. This means that data is not usually found in a tabulated and organized way. Therefore, it cannot be easily used for analysis and machine learning applications. This leads to the capture and selection of relevant information on a large scale cannot be done manually, instead, require certain degree of automation. It is at this point that natural language processing (NLP) comes into play.

In the last decades, there have been many initiatives to use NLP to mine data from unstructured documents and build usable databases for diverse fields. In particular, biomedical field can be highlighted to have numerous related datasets obtained through text mining[4]. For instance, Roche Diagnostics constructed a disease marker dataset upon extracting information from 50 million abstracts[5]. Other examples are the development of drug-drug interactions datasets and methodologies to extract drug side effects[6,7]. In turn, all these datasets have provided comprehensive access to the information and enable the use of artificial intelligence (AI) to boost the discovery of new molecules and improve therapies for patients[5,8,9]. In pharmaceutical manufacturing, most works building databases have focused on primary processing for the extraction of chemical reactions and materials[10–13]. In all these examples can be noticed how information extraction (IE), seen as an NLP task, provides a vast contribution in pharmaceutical data mining.

Information extraction (IE) aims to identify and summarise information of interest from documents of a specific domain[14]. This task is a multistep process that usually requires the deployment of several components. A first step usually involves preprocessing, which prepares text to be used downstream[15,16]. Nonetheless, preprocessing steps may also include methods to filter out irrelevant data to improve obtained information reliability[16]. Then, components like a part-of-speech (POS) tagger can be applied to individual words to assign semantic functions (verb, noun, or adjective). In a similar way, named entity recognition (NER) systems categorise words into a class, that hints at the type of data provided by each token[16,17]. Contrary to POS tagger, NER component is more domain-specific. Thus, the development of an IE system involves the design and elaboration of several components depending on the field of application.

In this regard, several workflows have been developed to extract information in the chemical domain. O. Kononova *et al* proposed a methodology to mine inorganic synthesis procedures from papers[18]. In this work, the authors developed methods that, firstly, identified paragraphs that contained the information on certain types of inorganic reactions. Then, these

EPSRC Future Manufacturing Research Hub for Continuous Manufacturing and Advanced Crystallisation (CMAC), University of Strathclyde, Technology and Innovation Centre, 99 George Street, Glasgow G1 1RD, UK.
E-mail: diego.alvarado-maldonado@strath.ac.uk

paragraphs in turn were input into other components to recognize materials, operations and conditions[18]. Continuing with primary manufacturing, Lowe's work developed a framework to extract chemical reactions and structures from patents[12], where the NER models are highlighted to play a key role. Similar to the aforementioned examples, many others have as a common factor the emphasis on the development of NER models[19–24].

NER systems can be considered the core of IE task[15]. This component identifies and enables the extraction of key information (entities). Depending on the methodology applied, NER can be based rules, or machine learning (ML) approaches[17]. Rule-based NER utilises dictionaries or semantic/syntactic patterns to label entities[17]. It has been shown that this approach offers very good precision but low recalls, due to the requirement of exhaustive dictionaries and the need for a wide variety of rules to cover all the possible variations in word patterns[17]. Owing to these limitations, ML methods have become more popular since they provide greater flexibility, achieving good recall and precision for several applications, particularly when deep learning is used[17]. In the pharmaceutical context, the development of NER models has been focused on biomedical and primary processing domain[10,12,13,19–23,25]. In the latter case, models for the recognition of materials, operations or both can be found. However, to the best of our knowledge, models for secondary processing have not been designed.

Getting back to IE workflows, pharmaceutical patents have special considerations to extract manufacturing data. Contrary to what happens in papers, patents are written in such a way that their procedures are difficult to reproduce and/or do not have well-defined sections which indicate whether or not a particular piece of information is available[11]. Moreover, the entire document does not necessarily cover a specific area, instead, involves several aspects of an invention that may include data of clinical studies, analytics, manufacturing procedures, etc. Therefore, the first step for data extraction should revolve around the selection of relevant sections or text fragments. Subsequently, these sections can be inputted into a NER model.

Nonetheless, as mentioned previously, current available NER models are mostly applied to primary processing data extraction for pharmaceutical manufacturing applications. Considering that secondary processing is out of the scope of these models, it would be needed to develop a model that enables to mine drug product fabrication data. An additional aspect to contemplate would be efficiency given a high volume of data. A unique model capable of performing the recognition of entities for both primary and secondary domains would allow extracting simultaneously all the targeted information. It would also facilitate distinguishing between synthesis/purification procedures and drug product manufacturing. In this manner, this work aims to explore and adapt tools that aid to the extraction of pharmaceutical manufacturing data of small molecules from patents. Understanding by pharmaceutical manufacturing, the fabrication of either active pharmaceutical ingredients (API) or drug products. In particular, this work focuses on two components 1) a relevant section selector and 2) a NER model for both primary and secondary processing using deep learning.

## 2. Methods

### 2.1. Collection of documents

Pharmaceutically relevant patents were identified, in first instance, by searching on the websites Bulk Search and Download API (uspto.gov) and PatentsView. As search terms, synonyms of drug substances obtained from drugbank database and dosage forms defined by Food Drug Administration (FDA) were employed[9,26]. There was a total of 43,538 search terms. From the search results, the patents with Cooperative Patent Classification (CPC) corresponding to the following groups were selected: A61K *Preparations for Medical, Dental, or Toilet Purposes*, A61P *Specific Therapeutic Activity of Chemical Compounds or Medicinal Preparations*, and A61Q *Specific Use of Cosmetics or Similar Toilet Preparations*. For those patents without CPC available, a deep neural network was used to identify whether these contained pharmaceutical data based on the abstracts. In the end, the detailed descriptions of the patents were retrieved from XML or TXT files, downloaded from United States Patent and Trademark Office (uspto.gov) website. This resulted in 208,596 patents being collected.

### 2.2. Relevant sections identification

#### 2.2.1. Text preprocessing

To discriminate those sections that contained specific information on manufacturing of small molecules, text clustering techniques were used. At this stage, the sections were identified based on headings, finding 5,542,816. The texts were then tokenized. Identified numbers were tagged as "[NUM]". Stop-words and punctuation were removed, and the remaining tokens were lemmatized. Tokenization was achieved using chemdataextractor 1.3, while the remaining steps were performed with spacy 3.3.

Afterwards, the sections were vectorised using a bag-of-words (BoW) representation. In this vectorization method, texts are represented by the token frequency considering a set of terms. For word counts, the terms that met the criteria shown in Table 1 were excluded. In the end, a vocabulary with approximately 62 thousand tokens was built. Subsequently, the sections with a number of tokens greater than $Q_{trunc}$ were truncated to this length. $Q_{trunc}$ was calculated making use of IQR rule for outliers, as indicated in Equation (1)[27], where $Q_1$ and $Q_3$ were the first and third quantile in text length distribution. This was done to ensure that very long texts did not affect clustering. The obtained representation was lastly used as an input in the next step.

$$Q_{trunc} = Q_3 + 1.5\,(Q_3 - Q_1) \qquad (1)$$

Table 1. Exclusion criteria for token in vocabulary for BoW representation.

| Criteria | Regular expression |
|---|---|
| Present in less than 200 hundred sections, | N/A |
| Present in more than 70% of all the sections, | N/A |
| Less than 3 characters, | N/A |
| Containing digits or punctuation, | "\d" and "(?u)\b\w\w+\b" |

### 2.2.2. Topic modelling

Latent Dirichlet Allocation (LDA) was used to categorize and represent sections based on latent topics. LDA is an unsupervised and probabilistic model that represents documents as a mixture of latent topics founded on word distribution[28]. This model has been widely applied for topic modelling and is particularly recommended when a large number of topics is suspected in the corpus (more than 20)[29]. LDA provides two major outputs, which are numeric representations for document-topics and topic-words[28,30]. The document representation is a vector where each element may be interpreted as the probability of membership to a certain latent topic or the contribution of each topic in document generation[29,30]. Regarding the topic-words distribution, the algorithm also assigns a probability to words depending on the topic, thereby being able to visualise the most likely words in each case[31]. Thus, this output can be used to associate latent topics with a particular subject by analysing the most important words in each.

In this manner, the preprocessed sections were split into 90% and 10% to train and test LDA models, respectively. Several LDA models were built varying the expected number of topics. In addition, models were run with and without shuffling and truncation of the training set. For every model, perplexity was calculated over the test set to select the optimal. Perplexity is defined as the inverse of the per-word likelihood geometric mean as shown in Equation (2)[28]. For a sample of $M$ documents, $\log p(\mathbf{w}_d)$ equals to per-word likelihood of the document $d$, and $N_d$ is the number of words in $d$. A lower perplexity is related to a model with a better generalization capability[28]. Thus, the model with the lowest perplexity for the test set was chosen as the best. The hyperparameters employed in the development of LDA models are summarised in Table 2.

$$Perplexity = \exp\left[-\frac{\Sigma_{d=1}^{M} \log p(\mathbf{w}_d)}{\Sigma_{d=1}^{M} N_d}\right] \quad (2)$$

Table 2. Hyperparameters assessed for LDA model development.

| Hyperparameter | Tested values |
|---|---|
| Alpha | Asymmetric |
| Iterations | 50 |
| Number of topics | 5, 10, 15, 20, 25, 30, 35, 40, 50, 60, 70, 80, 90, 100 |
| Passes | 1 |
| Chunk size | 4096 |
| Python library | Gensim 4.1 |

Moreover, the top 10 most important words per topic were determined for the best model. These words were firstly revised to identify meaningful associations within the topics and to validate how coherent the model outputs were. On the other hand, keywords were also employed to assign arbitrary topic labels for the cluster analysis which is detailed below.

### 2.2.3. Text clustering

LDA yielded the latent topics and their contributions for each section, by which the next step was to group documents that shared similar ideas. Therefore, to define clusters and allocate documents to a group, k-Means algorithm was employed, inputting the document representations obtained. k-Means is an iterative clustering technique. The algorithm works by generating k centroids, where k is a pre-set number of clusters[30]. Subsequently, the documents are assigned to a particular cluster based on the distance between documents and centroids, in such a way that every document is allocated to the cluster whose centroid is the closest[30]. Once all the documents have been assigned, centroids are recalculated by averaging document representations[30]. Subsequently, documents are reassigned using the updated centroids. This process is repeated until no significant changes in the centroids are observed or after a predefined number of iterations is completed[30].

In this work, due to the large number of data, a variation called minibatch k-Means was applied. In this version, instead of using all the data in every iteration, subsamples known as mini-batches are employed[32]. However, as in k-Means, the optimal number of clusters must be selected. Thus, various models were trained with k ranging from 5 to 60. To choose the best model, Davies-Bouldin (DB) and Silhouette (S) scores were calculated[33,34]. For the calculation of these metrics, random samples of 10 thousand documents were drawn. The calculations were replicated six times in different samples to estimate the variability of the scores. In the case of DB score, the optimal number of clusters is reached when the minimum value is found, whereas the opposite applies for S score [33,34].

As for the implementation of Minibatch k-Means, scikit learn library was employed. However, one of the limitations of this implementation is that only works with Euclidean distance. In this case, LDA document representations consists of probabilities. In this regard, Euclidean distance has been proven not to provide the best results for this type of data[30]. Alternatively, distances like cosine and Hellinger have been shown to measure document similarity better for clustering analysis [30,35]. Accordingly, data were previously transformed. As proxies of cosine and Hellinger distance, two transformations were assessed: L2-normalization and element-wise square root. The chosen distances are strongly related to Euclidean distance as illustrated in Equations (3) - (5)[35].

$$D_e^2(\mathbf{x}, \mathbf{y}) = \sum_{i=1}^{n}(x_i - y_i)^2 \quad (3)$$

$$D_h^2(\mathbf{p}, \mathbf{q}) = \sum_{i=1}^{n}\left(\sqrt{\frac{p_i}{2}} - \sqrt{\frac{q_i}{2}}\right)^2 \quad (4)$$

$$D_c(\mathbf{x},\mathbf{y}) = \frac{D_e^2(\mathbf{x},\mathbf{y})}{2}, if \ ||\mathbf{x}|| = 1 \ and \ ||\mathbf{y}|| = 1 \qquad (5)$$

Where $D_e, D_c, and \ D_h$ are Euclidean, cosine and Hellinger distances. $D_e = D_h$ if $x_i = \sqrt{p_i/2}$ and $y_i = \sqrt{q_i/2}$.

With the centroids of the best model, the latent topic with the highest contribution was selected in each cluster. Keywords obtained from LDA were thus revised, and a label that fitted with the keyword information was set. Afterwards, all the sections were assigned to a cluster. To validate the model, a random sample of 5 sections per cluster was drawn and each section was revised. If the section was clearly related to the arbitrary label, this was marked as 1 or, otherwise, 0. In the case there was no absolute certainty about the content, a value of 0.5 was given. At the same time, those sections whose content was more related to pharmaceutical manufacturing, for instance, having information on operations, methods, dosage forms or composition, were separately labelled with 1, whereas 0 was assigned if the documents did not meet this condition. Lastly, the agreement between the results of the algorithm and the manual assessment of relevant documents was estimated using Cohen's kappa and percentage of agreement. The former metric was used to measure agreement between manufacturing information and the latter to assess concordance between assigned label and actual section content. It is worth mentioning that two percentage of agreement were calculated, one for the worst case and another for the best. In the worst case, those sections with an assigned values of 0.5 were rounded to 0, while 1 was used for the other case.

### 2.3. Named entity recognition (NER)

#### 2.3.1. Training set preparation

A sample of 2069 paragraphs expected to contain relevant data were selected. These paragraphs were segmented into sentences and then tokenized using chemdataextractor v 1.3. Subsequently, using IOB scheme, every token was labelled manually using entity classes defined in ESI. IOB scheme is a label methodology employed in named entity recognition applications and stands for inside (I), outside (O), and beginning (B)[36]. This is particularly helpful to identify entities that are composed of more than one token. In this approach, irrelevant tokens are labelled as O. As for the remaining, the token that begins the entity is marked with B followed by the respective label. Then, the rest of the entity tokens are marked with I-[Label].

In the end, the training set consisted of 7440 sentences. As a final step, sentences with one token and duplicates were removed to obtain 7215 (221.257 tokens). These sentences were further corrected by comparing actual and predicted labels by the models and manually amending. The correction procedure was repeated twice randomising the examples order during training stage

Finally, the examples were split into training, development and test set, where sentences were distributed in the following percentages 80, 10 and 10, respectively. For the selection of the best model, the performance of NER model in the development set was monitored during training.

#### 2.3.2. Model training and validation

In the literature, various approaches for NER have been reported, however, with no doubts, methods based on deep learning (DL) have provided the best performance[17]. DL-based NER consists of a system of three components: an input representation, a contextualiser encoder and a label decoder[17]. The input representation generates a numerical depiction of each word or token. As representations, word embeddings such as word2vec, fasttext or transformers-based models are frequently used. In addition to word embeddings, other features can also be included such as character embeddings or part-of-speed information[17]. Then, the contextualiser processes information from inputs as a sequence, thereby considering the tokens order, to generate a contextual representation that feeds a decoder[17]. Typically, this component corresponds to bidirectional (Bi) recurrent neural networks with long short-term memory (LSTM) or gated recurrent unit (GRU) cells[17]. At last, the latter component translates inputs into entity types, assigning a label to each token. This task is normally achieved by using a conditional random field (CRF) layer[17]. This architecture is illustrated in Fig. 1.

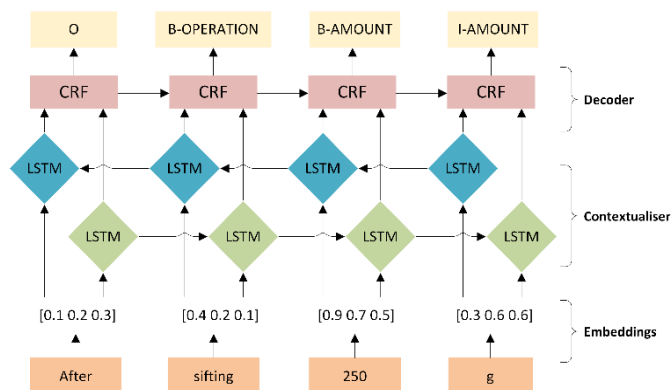

**Fig. 1** Common architecture for NER using DL.

As a base architecture in this work, fasttext embeddings-BiLSTM-CRF was employed. Word embeddings were pre-trained using gensim package with vector size and windows size of 300 and 5, respectively. As for the remaining hyperparameters, gensim default settings were used. For pretraining, all the collected descriptions were tokenized and used. This information fed a BiLSTM, whose results input a CRF layer. In addition, some variations were assessed which included the use of a multihead attention (MHA) layer after the embeddings. This latter layer was included as some works in this area have reported to provide good results[23,37]. Another was the application of convolutions after the embeddings (Conv1D). Finally, character embeddings (CE) were also evaluated, along with variations of dropout rates, as suggested elsewhere[36].

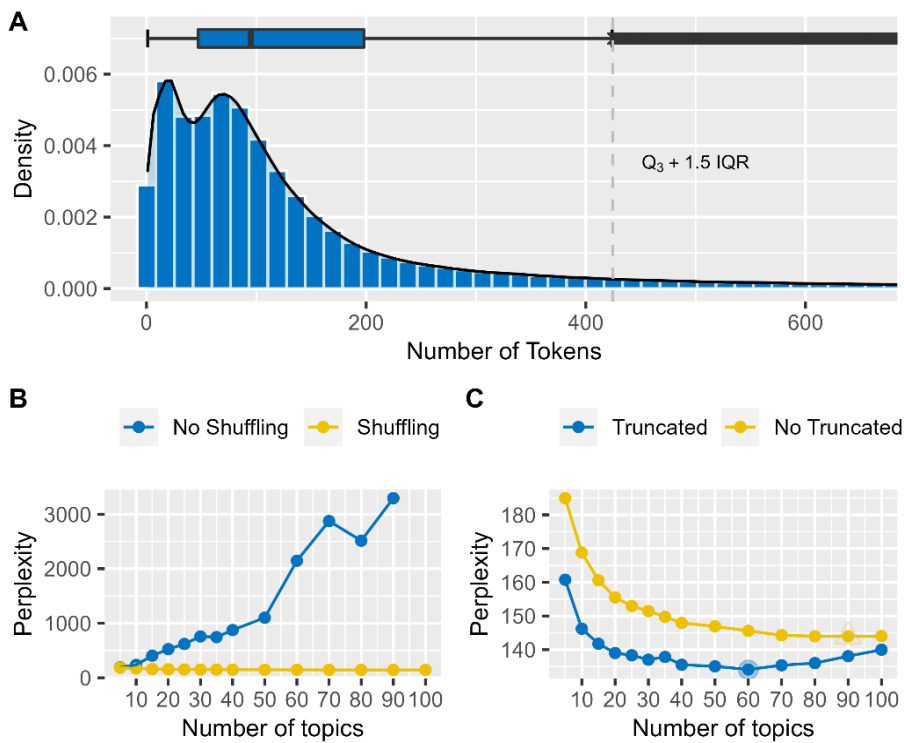

**Fig. 2** Distribution of the number of tokens per section (A) and Perplexities determined for LDA models evaluating shuffling (B) and truncation effect (C).

Deep neural networks were then trained using Tensorflow 2. AdamW was selected as the optimizer. As a performance measurement, f1-score micro average was utilised. The hyperparameters for the model were chosen based on preliminary experiments and these can be found in ESI. For training, early stopping strategy was used, where the training stopped when there was no improvement in the development set f1-score after 20 epochs. The best performing model was finally selected based on the highest f1-score micro average for both development and test set. With the chosen model, an error analysis was carried out to evaluate the most common mistakes

## 3. Results and discussion

### 3.1. Selection of relevant sections

#### 3.1.1. Optimal number of latent topics for LDA

LDA models were trained to determine the potential topics. During this process, the effect of examples order and inclusion of long texts were assessed. As for the former, it has been reported that LDA may suffer from "order" effect, which means that the order of the examples in the training set may affect the results[38]. This was confirmed in **Fig. 2**B where perplexity tended to increase every time the number of topics was greater. In contrast, when the order was randomized, the opposite behaviour was exhibited. Perplexity tends to decrease as a function of the number of topics, whereby shuffled data provided a behaviour that fits more with theory[28]. In this way, the results suggested that shuffling data is necessary to have reliable LDA models.

Regarding the effect of text length on model performance, the distribution for the number of words per preprocessed section and the effect of truncation can be seen in **Fig. 2**A and C. Nearly 87% of all the sections contained 425 tokens or fewer. This point corresponded to the superior limit of the boxplot, by which based on the IQR rule, values greater than this threshold. Thus, texts whose length was higher than $Q_{trunc}$ were defined as long. By observing **Fig. 2**C, truncated texts showed lower perplexities throughout the assessed range, compared to texts without truncation. This difference even led to a different optimal number of topics, with values of 60 and 90 for truncated and non-truncated models, respectively. Furthermore, apart from the difference in perplexity, the outputs quality was also affected, impacting on the topic interpretability as illustrated in **Fig. 3**.

Overall, when texts were not truncated, very specific words were found within the most important words. For instance, active ingredients or uncommon operations seemed to have a greater relevance (Topic 49 and 73 - **Fig. 3**B). Another aspect observed in the sections without truncation was that the LDA model yielded topics with keywords that were difficult to interpret more often (**Fig. 3**B – Topic 23). Lastly, whereas there were topics with a distribution which clearly favour certain words, some groups had keywords with more uniform probabilities (**Fig. 3** – Topics 32 and 73). In consequence, it is possible that these keywords did not generalise well the topics, by which the documents in these groups might not necessarily be related to what keywords suggested, reducing model interpretability and reliability. Once again, although this happened in both models, these kinds of topics were less frequent for truncated data. In this manner, the truncated model was concluded to be the best, with 60 latent topics. Furthermore, section truncation was also demonstrated to generate more understandable topics. Keywords for both models can be found in Appendices A and B.

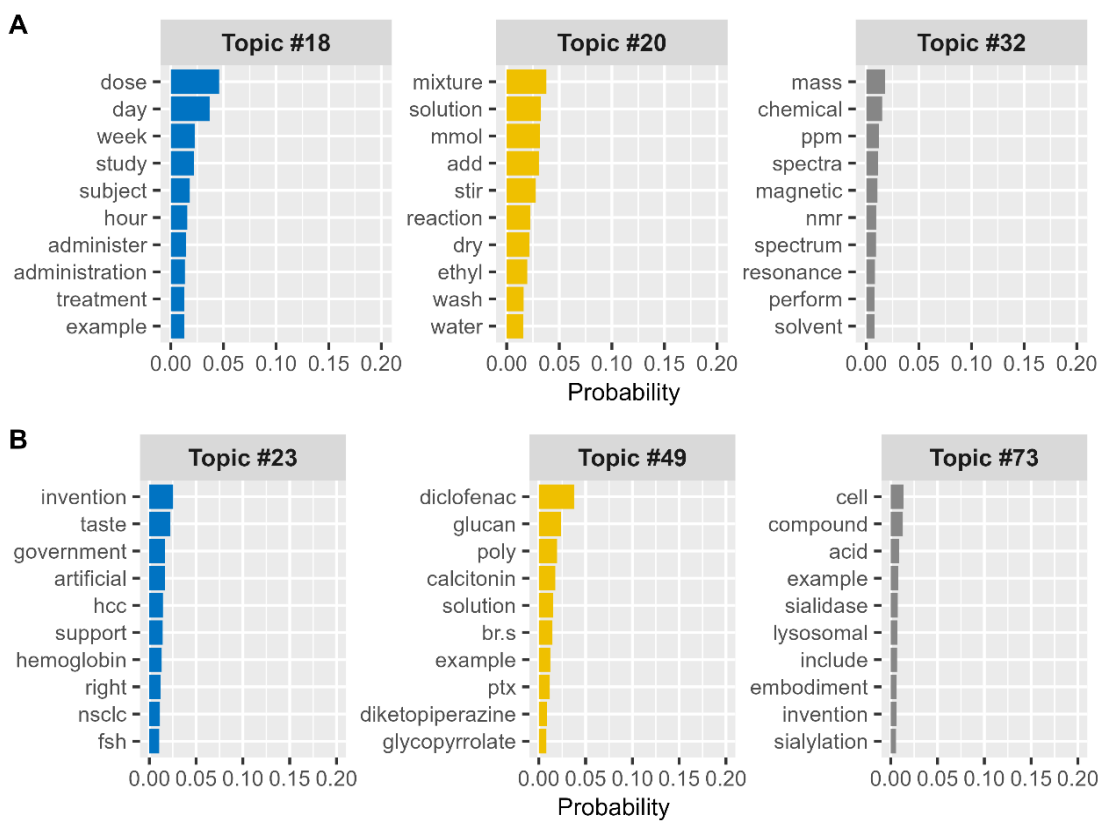

**Fig. 3** Top 10 keywords for truncated (A) and non-truncated (B) LDA models.

#### 3.1.2. Optimal number of clusters for k-Means

The results of the scores for the selection of the optimal number of clusters can be found in ESI. The first factor analysed was the effect of data transformation. Across the assessed range and for both metrics, L2-norm outperformed square root transformation. Knowing this, the analysis next focused on selecting the optimal number of clusters for normalised data. The optimal points given by DB and S scores differed. However, one of the main differences between both scores was the variability. S score showed values with a higher scattering compared to DB score. Therefore, the differences in S score might have not been as significant as with DB index. Due to this consideration, the selected number of clusters for k-Means was 60, which was the optimal considering DB index.

#### 3.1.3. Algorithm performance and outputs interpretability

**Fig. 4** depicts LDA representation for a sample of 2000 sections, using t-distributed stochastic neighbour embedding (t-SNE) to reduce dimensionality. A total of 17 labels were assigned to the clusters. It was possible to visualise how documents belonging to the same cluster and label tend to be close to each other. An example of this can be seen in the references 1 and 2 (black points) in **Fig. 4**, whose texts can be seen in **Fig. 5**. These two sections were about to synthesis procedures, and it can be seen how the algorithm correctly classified them into a primary manufacturing topic, having similar values. In the same manner, upon selecting labels related to manufacturing, LDA representations allowed to agglomerate documents into two clearly defined regions about pharmaceutical manufacturing. These results supported that LDA enabled to compare numerically documents which are semantically similar. In addition, LDA + k-Means enabled to separate information into interpretable topics, considering this work scope.

The results of performance assessment are found in ESI. In this case, the agreement between sections and labels assigned can be seen for the best and worst case. The overall percentage of agreement in the worst case was around 81%. The lowest degree of agreement was found for the cluster assigned with

| Ref. 1 (Patent Number US10590109B2) | Ref. 2 (Patent Number US8324225B2) |
|---|---|
| **Step 9** 4-((2-(dimethoxymethyl)-6-(methylamino) pyrid-3-yl)methyl) morpholin-3-one<br><br>Compounds t-butyl-(6-(2-methoxyethyl)-5-((3-carbonylmorpholine)methyl) pyrid-2-yl)(methyl) aminocarboxylate 9i (70 mg, 0.18 mmol), trifluoroacetic acid (1 mL) and dichloromethane (4 mL) were mixed, and stirred for 6 h at room temperature. The mixture was alkalified with triethyl amine, and subjected to exsolution under reduced pressure. The residuals were purified through a preparative silica gel plate (petroleum ether/ethyl acetate 1:1), to obtain the target product 4-((2-(dimethoxymethyl)-6-(methylamino) pyrid-3-yl) methyl) morpholin-3-one 9j (46 mg, colorless solid), at a yield of 86%.<br><br>MS m/z (ESI): 296 [M+1]. | **EXAMPLE 377** 1-[2-[4-(4-Acetyl-piperazin-1-yl)-phenylamino]-7-(1-ethyl-propyl)-7H-pyrrolo[2,3-d]pyrimidin-6-yl]-ethanone<br><br>To a solution of 1-[2-chloro-7-(1-ethyl-propyl)-7H-pyrrolo[2,3-d]pyrimidin-6-yl]-ethanol (61 mg, 0.2 mmol) in $CH_2Cl_2$ (2 mL) is added Dess-Martin periodinane (242 mg, 0.5 mmol). The reaction mixture is stirred for 1 h, quenched with 10% $NaS_2O_3$:saturated $NaHCO_3$ (1:1) aqueous solution, and extracted with $CH_2Cl_2$. The extracts are washed with water and brine, dried over $Na_2SO_4$, and concentrated in vacuo. The residue is purified by flash chromatography ($SiO_2$, EtOAc/Hexane 1:3) to afford 58 mg of 1-[2-chloro-7-(1-ethyl-propyl)-7H-pyrrolo[2,3-d]pyrimidin-6-yl]-ethanone.<br><br>LCMS: 266 (M+H)$^+$. |

**Fig. 5** Example documents with similar content and classified in the same cluster (primary manufacturing) referenced in **Fig. 4**.

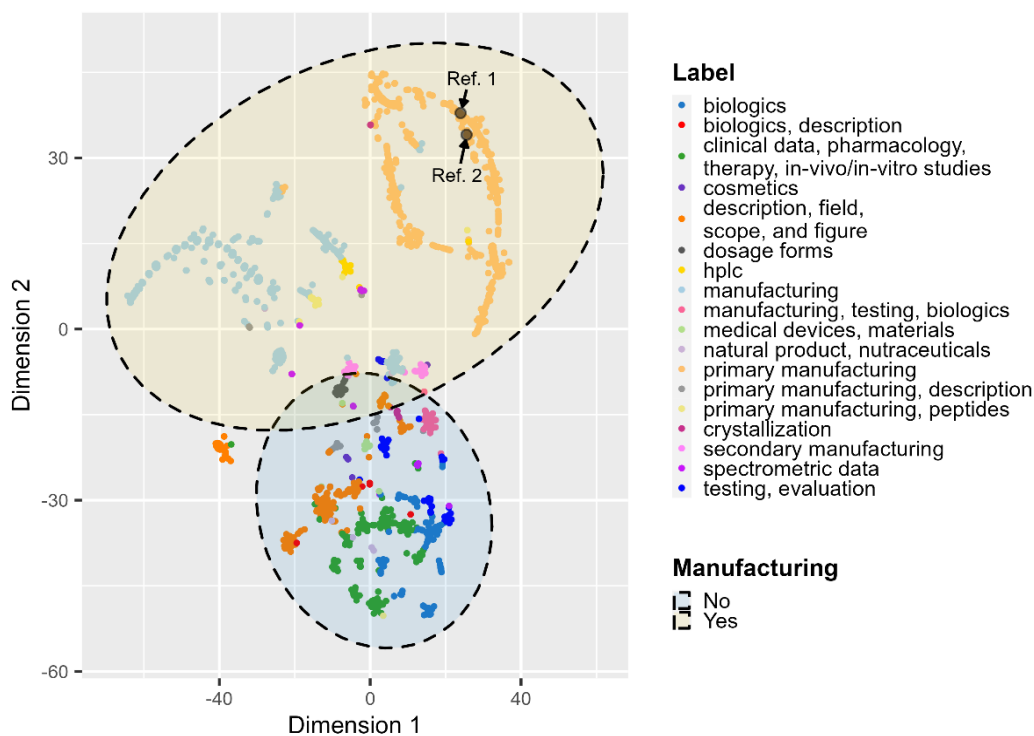

**Fig. 4** t-SNE visualization of document representations generated using LDA.

spectrometric data label. The LDA topic most important for this label corresponded to the number 32, which was characterised for having keywords with uniform probabilities, as shown in **Fig. 3**A. As discussed previously, this fact results in keywords with a poor power of characterisation, that results in low generalization. As a result, these keywords were not helpful to define a reliable label. Another unique aspect of the documents classified into this label was that LDA representations were widely spread with respect to the others. This could be evidenced in **Fig. 4** where documents belonging to analytical topics were very sparse, and it was difficult to appreciate a dense cluster. In this manner, this cluster in particular was not reliable and grouped sections non-related with the assigned label. Nonetheless, despite this cluster behaviour, it is important to highlight the cluster did not contain sections on manufacturing considering the revised sample. Likewise, the label was assigned as non-relevant since it is focused on spectrometric results. Therefore, for manufacturing information retrieval, the sections belonging to this cluster were not considered for the subsequent data extraction step, whereby this should not affect final results.

The remaining labels presented values greater than or equal to 50%, leading to an overall performance greater than 80% for both cases. This suggests that there is a high agreement between the information provided by the documents and the interpretation provided by the topic keywords and clusters, at least being better than a random guess. Given there was a good level of agreement between the cluster interpretation and the grouped sections, it was possible to generalise to a higher level by relating directly the labels with the possibility of containing information on manufacturing, with a special emphasis on small molecules. It was thus all the clusters with a label about manufacturing was marked as relevant. The list of labels with relevant and irrelevant information can be seen in **Error! R** eference source not found.. In this manner, this methodology enabled to indirectly determine what sections were of interest.

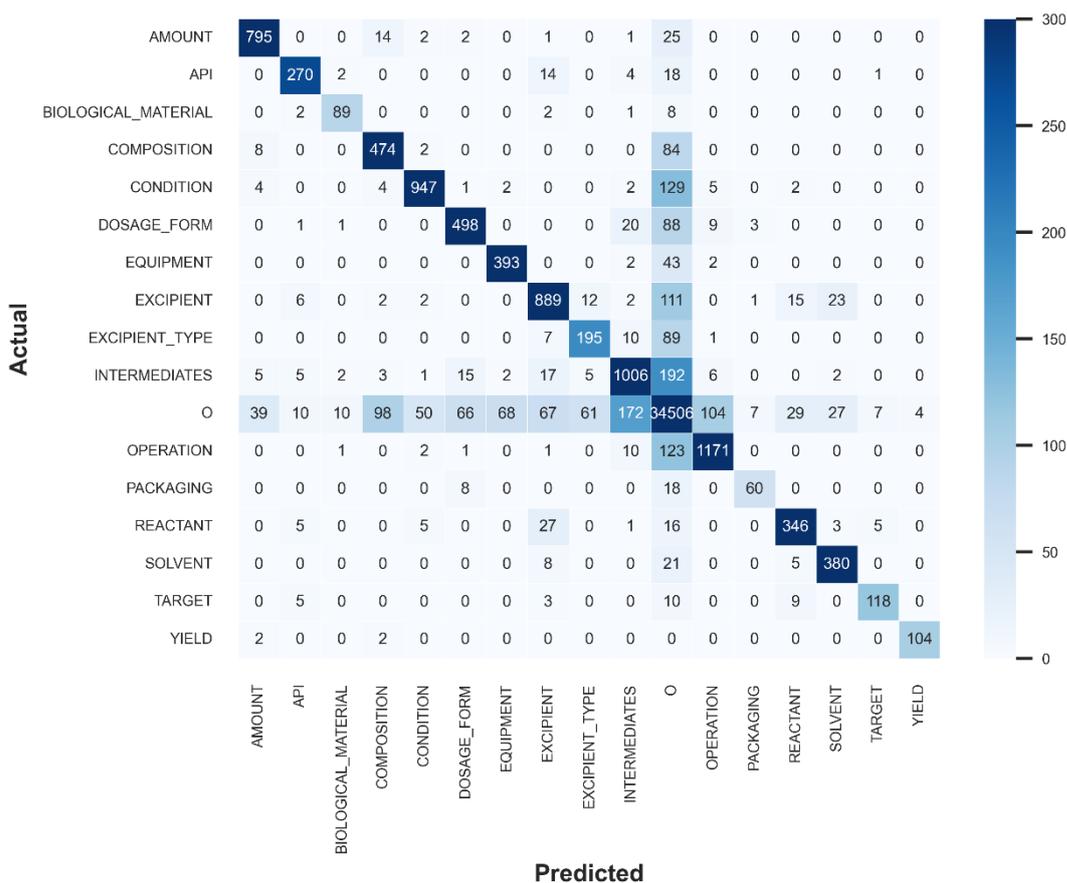

Fig. 6. Confusion matrix for NER

In the end, 22 out of 60 clusters were considered relevant. Although, there were a few labels considered irrelevant that still mentioned aspects about fabrication and composition, these were discarded as pharmaceutically relevant because they only provided generic information, for instance, some sections listed all the possible dosage forms. Interestingly, when the agreement was calculated by relevancy, the group of manufacturing had a greater level with 88.1% for the worse case, whereas 77.8% was obtained for the other group. Some possible reason behind that were the presence of few sections that were related to manufacturing, and perhaps the most frequent, sections that could be associated with other similar labels. Finally, with the model output and manual revision, agreement was measured using Cohen's kappa for manufacturing data. This resulted in a value of 91.1%, which can be considered an acceptable[39]. In this way, the results suggested that the algorithm was able to distinguish sections with information of interest from those which were irrelevant. Nonetheless, it is worth to highlight that even though this approach may help to distinguish relevant information in an efficient manner, there still sections in the sample which were difficult to define thereby the final selection might still contain a reduced number of irrelevant information.

### 3.2. NER

#### 3.2.1. Selection of Best architecture

The summary of the results can be seen in Table 3. Using as a base model the architecture without convolution or attention layers (BiLSTM), it can be seen that the self-attention layer did not enhance f1-score in any case. Instead, a reduction of 2.6% on average when used was observed. On the other hand, Conv1D seems to have a significant improvement in model performance, helping to filter relevant semantic features[40]. A similar effect was expected for the attention layer; however, this was not the case. In this manner, for the best model, the attention layer was discarded, whereas the Conv1D layer was demonstrated to be needed to obtain the best possible performance.

In addition to the inclusion or exclusion of specific layers, the confirmatory test also comprised the evaluation of the use of additional features like characters embeddings (CE). CE have proven to enhance NER performance in several domains[36]. To some extent, this is achieved by modelling other features not considered by word embeddings such as word morphology and spelling[36]. However, the effect of CE might be obscured by the effect of word features[36]. Due to this, some authors recommend adjusting dropout to maximise the effect of CE and model performance. Therefore, tests using CE with several levels of dropout were also assessed[36]. As observed in Table 3, the additional features increased model performance independent of the rates of dropout used, with f1-scores surpassing 83.0% in all cases. Upon comparing the behaviour of performance metrics using different dropouts, a disparity in trends were observed. For the development set, the higher the dropout, the better the performance, while the test set resulted in the opposite. Nonetheless, on average, the former trend was dominant, whereby the model with the highest dropout was selected as the best. The best model architecture is illustrated in ESI.

Table 3. F1-scores (micro-average) for confirmatory models in development and test sets. BiLSTM: Bidirectional Long Short-Term Memory, Conv1D: unidimensional convolution, Attention: multihead self-attention, CE: Character embeddings.

| Model | Dev | Test | Average |
|---|---|---|---|
| BiLSTM | 81.7% | 79.6% | 80.6% |
| BiLSTM + Conv1D | 84.6% | 82.0% | 83.3% |
| BiLSTM + Attention | 79.4% | 76.7% | 78.0% |
| BiLSTM + Attention + Conv1D | 83.8% | 82.0% | 82.9% |
| BiLSTM + Conv1D + CE + Dropout (0.3) | 84.7% | 83.5% | 84.1% |
| BiLSTM + Conv1D + CE + Dropout (0.5) | 84.9% | 83.2% | 84.0% |
| BiLSTM + Conv1D + CE + Dropout (0.7) | 85.4% | 83.0% | 84.2% |

### 3.2.2. Recognition performance and error analysis

The performance of the final model for each entity is broken down in Table 4. The overall performance, in terms of precision and recall, was 84.9% and 83.5%. These values were comparable to other works in similar domains, where results revolved around 60 and 98%[16]. To highlight, the top 3 most difficult entities to recognize were related to packaging materials, excipient type, and targets, having the lowest values for the different metrics. A key aspect of these types of entities is the low recall, which is translated into a loss of information during extraction. From the confusion matrix shown in Fig. 6, the most common error for these entities was not to be recognized or classified as irrelevant ("O"). In addition, this error was also the most frequent among the other types of entities.

Interestingly, in the specific cases of material-related entities such as targets or packaging, these were also misclassified into another material class such as reactant or dosage form, respectively. It was noticed that this behaviour could be extrapolated to any other similar type of entity, for instance, API, excipients or solvents. Consequently, it can be concluded the model was associating the materials with material-concept tags. Going to other types of entities such as amount, composition and yield, these tended to be classified into one another, when misclassified. However, despite the error, the model was still associating entities with similar concepts. In contrast, this trend was not observed in operations or conditions. Operations were sometimes tagged as intermediates or dosage forms were classified as operations. Nonetheless, these cases were marginal compared to operations falling into "O" class, representing 0.8% and 1.5% of the predictions, respectively. Furthermore, other types of errors were even less frequent. Thus, overall, the model yielded errors which could be considered reasonable in most cases. However, for information retrieval and machine learning development, further processing and cleaning steps may be required to mitigate the effect of these errors on extracted data reliability. As for the possible causes of these errors, these might be explained by two sources: context modelling and training set size.

Table 4. Breakdown best model performance.

| Entity | Precision | Recall | F1-Score |
|---|---|---|---|
| AMOUNT | 94.1% | 95.0% | 94.5% |
| API | 88.1% | 88.2% | 88.1% |
| BIOLOGICAL_MATERIAL | 87.2% | 83.6% | 85.3% |
| COMPOSITION | 75.8% | 81.1% | 78.1% |
| CONDITION | 89.3% | 83.3% | 86.2% |
| DOSAGE_FORM | 80.7% | 82.4% | 81.5% |
| EQUIPMENT | 80.2% | 81.3% | 80.6% |
| EXCIPIENT | 83.3% | 79.1% | 81.1% |
| EXCIPIENT_TYPE | 70.9% | 64.6% | 67.6% |
| INTERMEDIATES | 78.5% | 75.5% | 76.9% |
| OPERATION | 89.6% | 89.1% | 89.4% |
| PACKAGING | 66.5% | 58.0% | 61.8% |
| REACTANT | 81.2% | 83.4% | 82.3% |
| SOLVENT | 88.7% | 88.6% | 88.7% |
| TARGET | 73.2% | 72.9% | 72.6% |
| YIELD | 97.4% | 94.4% | 95.7% |
| micro avg | 84.9% | 83.5% | 84.2% |

In the first instance, sentences were used as inputs for NER model and, for this type of input, the surrounding words might have not sufficed to know whether or not a token belongs to a specific category, occasionally. An example of this was seen with materials which can be used as excipient or reactant like Sodium Hydroxide or Hydrochloride Acid, where in a synthesis procedure, the model recognized them as EXCIPIENT. Another possible reason may be that the employed embeddings are not capturing completely the differences between the entities. In this regard, the embeddings deployed for the present work generate a unique representation for each token. Recent developments have provided embeddings based on transformers that generate representations for tokens considering the context[10,16]. In this way, the same word may have several representations depending on the other surrounding words. This approach has shown good results in other domains [10,16,41], whereby can be worth exploring as an opportunity for improvement in future works.

Going back to the cases of targets and packaging materials, a characteristic of these types of entities was that there were not many examples in the training set. Nearly 2.0% of the sentences had terms related to packaging materials, while around 2.6% contained tokens about targets. In this manner, the inclusion of examples containing this type of entity with a greater diversity of terms might help the model learn to identify these better. Although it was observed that these entities were not mentioned very often throughout the patents compared to not being identified. In particular, the main focus in the studied patents was formulations, whereby packaging materials were not usually mentioned. In the case of targets, a common pattern in the analysed document was to reference implicitly and chemical structure or name was described in different sections. Thus, these entities are not expected to be abundant in the corpus, hindering found more training data.

## 4. Conclusions

In this work, a set of tools to aid information extraction of primary and secondary manufacturing from pharmaceutical patents were developed. The first component comprises a sections selector. This model filters sections that potentially contains information on manufacturing, with special emphasis on small molecules. In the end, LDA + k-Means method grouped documents, and provided good interpretability for the majority of clusters based on keywords. The validation yielded an overall agreement between assigned topic labels based on cluster keywords and the actual section topic was 81.7%, in the worst case. For the identification of manufacturing relevant documents, the agreement was measured using Cohen's kappa, obtaining a value of 91.1% between the outputs of the model and the actual section topic, which can be considered a good level of agreement. This suggests the applied approach has can be useful to select relevant fragments of text.

On the other hand, once the fragments of text containing the relevant information have been filtered, it is necessary to apply another algorithm that recognizes the specific information we are interested in extracting. To accomplish this, a NER model was trained. This model consisted of a deep neural network with a base architecture comprising word embeddings, a contextualiser (BiLSTM) and a decoder (CRF). The overall performance for the NER model, measured using F1-score micro average, was 84.2%. The most common errors were associated with the inability to detect certain entities, followed by the confusion between recognized entities, to a lesser extent. For instance, a material which is a final product of a synthesis (TARGET) could be confused with a reactant. However, these types of errors seem to be occasional. Future work focused on improving the NER algorithm may focus on using word embeddings based on transformers. However, the developed model showed to be able to recognize the required information with a level of accuracy comparable to analogous works. Finally, another aspect to be covered in future works is the integration and application of the developed tools to build a dataset for secondary manufacturing processes. Some consideration to bear in mind when applying these tools is the need of post-processing to mitigate the effect of errors on extraction reliability.

## Data Availability Statement

All data underpinning this publication are openly available at https://github.com/Diego-Alvarado/nlp_for_pharma_manufacturing.git:
- Preprocessed patents for unsupervised model development (LDA+k-Means).
- Annotated sentences for pharmaceutical manufacturing NER.
- Scripts for the training of the model.
- Sections used to validate LDA+k-Means.

## Conflicts of interest

The authors declare no competing financial interest.

## Acknowledgements

The authors would like to thank EPSRC and the Future Continuous Manufacturing and Advanced Crystallisation Research Hub (Grant Ref: EP/P006965/1) for funding this work.